\documentstyle[preprint,aps]{revtex}

\begin{document}
\draft
%\preprint
%\narrowtext
%\twocolumn

\renewcommand{\theequation}{\arabic{equation}}
\renewcommand{\thesection}{\arabic{section}}

\title{Theory of a single Oxygen hole propagating in ${\rm Sr_2CuO_2Cl_2}$:\\
the spin of the quasiparticles in ${\rm CuO_2}$ planes}

\author{K. J. E. Vos} 
\address{Department of Physics, Queen's University,\\ Kingston, 
Ontario, Canada}

\author{R. J. Gooding\footnote{Permanent address: Dept. of Physics, Queen's
University, Kingston, Ontario, Canada.}}
\address{Department of Physics and the Center for Materials Science 
and Engineering,\\ Massachusetts Institute of Technology, Cambridge,
Massachusetts, U.S.A.}

\date{\today}

\maketitle 
\vskip 3.0 truecm
\pacs{Submitted to Z. Phys. B}
\newpage
\begin{abstract}

Recent photoemission experiments have measured $E~vs.~\vec k$ for a single
hole propagating in antiferromagnetically aligned ${\rm Sr_2CuO_2Cl_2}$. 
Comparisons with (i) the $t -t' - J$ model, for which the carrier is a spinless
vacancy, and (ii) a strong--coupling version of the three--band Emery model,
for which the carrier is a $S = {1\over 2}$ hole moving on the Oxygen
sublattice, have demonstrated that if one wishes to describe the quasiparticle
throughout the {\em entire} first Brillouin zone the three--band
model is superior. Here we present a new variational wave function
for a single Oxygen hole in the three--band model: it
utilizes a classical representation of the antiferromagnetically ordered 
Cu--spin background {\em but} explicitly includes the quantum fluctuations 
of the lowest energy doublet of the Cu--O--Cu bond containing the 
Oxygen hole. We find that 
this wave function leads to a quasiparticle dispersion for physical exchange 
and hopping parameters that is in excellent agreement with the measured ARPES 
data. We also obtain the average spin of the Oxygen hole,
and thus deduce that this spin is only quenched to zero at certain wave
vectors, helping to explain the inadequacy of the $t - t' -J$ model
to match the experimentally observed dispersion relation everywhere in 
the first Brillouin zone.

\end{abstract}
\vskip 2.0 truecm
\pacs{Keywords: Strongly correlated electronic systems; high $T_c$ ${\rm CuO_2}$
planes;\hfill\break 3--band Hubbard model}
\newpage

% ** Introduction (including a full discussion of the Hamiltonian) **
% ----------------
\section{Introduction}
\label{sec:introduction}

The high $T_c$ transition metal oxides have been intensely studied, with
the focus of most researchers directed towards the physics of the 
CuO$_2$ planes. Undoped, these planes are composed of 
S = ${1\over 2}$ spins at each Cu site, and it has been shown that
a simple Heisenberg Hamiltonian with a near--neighbour exchange of
the order of 1500 K provides an excellent representation of the
antiferromagnetic (AFM) interactions between the spins over a
wide temperature range \cite {greven}.

Having confidence in a good model of the undoped state,
a theory must be developed dealing with the motion of 
carriers in a Cu--spin background.  
Upon doping the planes mobile holes are produced,
and these holes are known to predominantly occupy the Oxygen sites in 
the CuO$_{2}$ planes. However, Anderson first proposed that one--band
models are sufficient to represent the important low--energy 
physics of such carrier motion \cite {anderson}.  
This proposition was formalized in work by 
Zhang and Rice \cite {zhang_rice}, and later by Shastry \cite {shastry}: 
starting from a three--band Hubbard model they delineated how it is possible 
to map the Oxygen hole motion onto a one--band $t - t^\prime - J$ model
(in fact, in Shastry's formulation, hopping processes connecting
sites that are further than next--nearest--neighbour are allowed).
It is to be emphasized that at the end of this renormalization, the
spin degree of freedom of the Oxygen carrier described by the
three--band Hubbard model is quenched, and the carrier described
by the $t - t^\prime - J$ model is a {\em spinless} vacancy.
While this renormalization has been the subject
of some debate, no definitive experimental data have allowed for
one to say that the one--band model is indeed a viable
representation of carrier motion. Instead, one--band model researchers 
are forced to adopt something similar to the following posture: the physics of
the $t - t^\prime - J$ model is sufficiently difficult to warrant
the attention of a high $T_c$ theorist, and hopefully
the physics of the superconducting instability and the
anomalous normal state will be contained in this model.

Fortunately, recent experimental work does allow for careful
qualitative and quantitative scrutiny of this assumption. 
To be specific, angle--resolved photoemission spectroscopy 
(ARPES) of insulating ${\rm Sr_2CuO_2Cl_2}$ was recently 
completed \cite {wells}. This experiment presents
theorists with, for the first time, data for the spectral function
of a {\em single} hole propagating in a ${\rm CuO_2}$ plane. In fact,
${\rm Sr_2CuO_2Cl_2}$ is an ideal model system to obtain such data 
since complications associated with the orthorhombicity
of many high $T_c$ materials can be avoided --- ${\rm Sr_2CuO_2Cl_2}$
does not undergo such a structural phase transition until about 10 K
and reliable ARPES data were collected at 350 K.

Subsequent to this work, a comparison of the $t - t^\prime - J$
model's $E~vs.~\vec k$ results to the ARPES data of Ref. \cite {wells}
displayed that the one--band model did not reproduce the
bandstructure everywhere in the AFM Brillouin zone
\cite {nazarenko}. Instead, in this same paper \cite {nazarenko}, 
we summarized our results showing that the three--band model 
in the strong--coupling limit
did a superb job of reproducing the $E~vs.\vec k$ ARPES
data for all wave vectors. Here we elaborate on that work and also display
a new, more detailed comparison to experiment.

In this paper we focus on a model of Oxygen holes moving
in a ${\rm CuO_2}$ plane. A new variational wave function for this
problem is introduced and applied to the ${\rm Sr_2CuO_2Cl_2}$ ARPES
data. This variational wave function is similar in spirit to
the successful semi--classical variational wave functions introduced by
Shraiman and Siggia for the one--band model \cite {SS} and later
generalized to the three--band model \cite {frenkel}. However,
unlike the variational wave function of Ref. \cite {frenkel},
here we include the lowest--order quantum fluctuations of the
occupied Cu--O--Cu bond, thus allowing for the so--called
trion quasiparticle \cite {ER} to be accounted for. These quantum
fluctuations permit, in principle, for the spin of the Oxygen
hole to be quenched to zero, an essential ingredient in the Zhang--Rice 
renormalization \cite {zhang_rice} of the three--band Hubbard model to the
one--band $t- t^\prime - J$ model. 

We show that our new variational wave function for a single Oxygen hole provides
an excellent representation of the $E~vs.~\vec k$ relation
measured in Ref. \cite {wells} for physically reasonable
exchange and hopping parameters. We then analyze these
wave functions and show that unlike the result of the renormalization of 
the three--band to the one--band model, the spin of the carrier
is not quenched to zero everywhere in the Brillouin zone.
This thus provides us with an understanding of the inability 
of the $t - t^\prime - J$ model to reproduce the
$E~vs.~\vec k$ data at all wave vectors \cite {nazarenko}.
Further analysis has allowed us to reduce this quasiparticle
to that of a doublet (the ground state of
three S=${1\over 2}$ spins antiferromagnetically coupled,
as in the occupied Cu--O--Cu bond) moving in a distorted
N\'eel background of Cu spins, {\em viz.}, the trion quasiparticle
of Ref. \cite {ER}. 

Our paper is organized as follows. In \S 2 we describe the components
of the three--band model in the strong--coupling limit.
In \S 3 we present our new variational wave function,
a description particular to the three--band model and one
that contains some quantum fluctuations of the propagating
Oxygen hole. This section also contains a comparison of
our results to the ARPES data of Ref. \cite {wells}.
Section 4 presents an analysis of the wave function that results
from the fit to the ARPES data, including the nature and spin of
the quasiparticle.  Finally, \S 5 summarizes our main findings
and describes future work on this problem.

% ** Model Hamiltonian 
% ----------------
\section{Components of the Three--Band Hamiltonian}
\label{sec:3band model}

The application of the extended three--band Hubbard model to a description
of the ${\rm CuO_2}$ planes 
of the high $T_c$ oxides was first discussed by Emery \cite {emery} and 
Varma {\em et al} \cite {varma}.  The parameters that are believed to enter this
model have been evaluated elsewhere \cite {stechel,hybertsen}. Here
we simply note that if one takes a ${\rm 3d^{10}2p^6}$ configuration
as the vacuum state, then in the strong--coupling limit one can derive
the following Hamiltonian as representative of the motion of Oxygen holes
in a Cu--spin background \cite {frenkel,ER}:
\begin{eqnarray}
{\mathaccent 94 H}& = &
J_{Cu-Cu} {\sum_{<i,i^{\prime}>}}^{\prime} \Big( \vec{S}_{i} \cdot
\vec{S}_{i^{\prime}} - 1/4 \Big)
+ J_{Cu-O} \sum_{<i,l>} \Big( \vec{S}_{i} \cdot \vec{S}_{l} - {1 \over 4} n_{l}
\Big)
\nonumber \\ &   &
+ (t_{a} + t_{b}) \sum_{<l^{\prime}il>,\sigma,\sigma^{\prime}}  \Big(
\vec{S}_{i} \cdot \Big( b_{l^{\prime},\sigma^{\prime}}^\dagger
{\mathaccent 94 \tau}_{\sigma^{\prime},\sigma} b_{l,\sigma} \Big) + h.c. \Big)
\nonumber \\ &   &
+ {1 \over 2} (t_{b} - t_{a}) \sum_{<l^{\prime}il>,\sigma} \Big(
b_{l^{\prime},\sigma}^\dagger b_{l,\sigma} + h.c. \Big)
- \big| t_{pp} \big| \sum_{<l^{\prime}l>,\sigma} \Big(
b_{l^{\prime},\sigma}^\dagger b_{l,\sigma} + h.c. \Big).
\label{eq:3tJ}
\end{eqnarray}
\noindent In this equation $i$ refers to Copper sites and $l$ refers to Oxygen
sites; the $\vec S_i$ are the Cu spin operators, whereas the spin operator 
for an Oxygen hole is ${\vec{S}_l} = {1 \over 2} b_{l,\sigma}^\dagger
\vec{\tau}_{\sigma,\sigma^{\prime}} b_{l,\sigma^{\prime}}$, where
$\vec{\tau}_{\sigma,\sigma^{\prime}}$ are the Pauli matrices,
and $b_{\ell,\sigma}$ is the destruction operator for an Oxygen hole
that precludes double occupancy. The primed summation indicates that
the superexchange is not calculated for the bond containing the 
Oxygen hole. The $t_a$ hopping process corresponds to the Oxygen hole passing
through a Cu site without affecting the Cu hole, whereas the $t_b$ hopping
exchanges the carrier with the hole of a Cu site ---
a more detailed description of the hopping 
processes and notation for this Hamiltonian are provided in
Ref. \cite {frenkel}. Implicit in Eq.~(\ref{eq:3tJ}) is the reduction
of all Oxygen $p$ and Cu $d$ orbitals to one's having an effective s--wave
character --- {\em e.g.}, see Ref. \cite {ER}.

The Cu--Cu superexchange parameter has been measured to be 0.125 eV
\cite{greven}, and here we fix $J_{Cu-Cu} = 0.125$ eV. Further, it has been
argued \cite{frenkel,ER} that $J_{Cu-O}/J_{Cu-Cu} \sim 4 \rightarrow 6$, and
thus we fix $J_{Cu-O} = 0.63$ eV \cite {630}. To estimate the hopping energies,
perturbation theory \cite {frenkel,ER} and electronic structure work
\cite{stechel,hybertsen} may be used. (In the perturbation theory we have used
a reduced $pd$ energy overlap integral $t_{pd} = 1.25$ eV to account, in a
first--order fashion, for the larger lattice constant of ${\rm Sr_2CuO_2Cl_2}$
-- this value, when employed with perturbation theory, accurately predicts the
Cu--Cu superexchange.) We find
\begin{equation}
t_a \sim 0.32~{\rm eV},~t_b \sim 0.36~{\rm eV},~{\rm and}~t_{pp} \sim
-0.65~{\rm eV}.
\label{eq:hybert 3tJ energies}
\end{equation}
These numbers are similar to other published sets of estimates.

We have chosen to fix both the experimentally determined $J_{Cu-Cu}$,
as well as our value of $J_{Cu-O}$ = 0.63 eV (varying $J_{Cu-O}$ by
$\pm 25$ per cent does not affect our results). However, the other 
parameters we allow
to become fitting parameters, and we choose them by fitting to the $E~ vs. \vec k$~ARPES data. 
Of course, if our fitted parameters are too far from those expressed in
Eq.~(\ref{eq:hybert 3tJ energies}) then it is not at all clear that
our fits are in any way physically reasonable. Fortunately, we find reasonably close
agreement with these numbers.

% ** New Variational Wave Function **
% ----------------
\section{New Variational Wave Function}
\label{sec:wave_function}

One successful approach \cite {SS} to the problem of a vacancy moving in an AFM background
described by the $t - J$ model was proposed by Shraiman and Siggia. In 
their formalism the undoped state corresponds to a classical AFM, a state which 
has as its wave function a simple product of spinors. Then they suggested
that one way to determine the symmetry and spin texture of the singly--doped ground state 
involved taking a semiclassical variational wave function based on a product state of spinors 
representing (again) a classical Cu--spin background  that included a single vacancy. Bloch states
can be formed from this product state, and then the direction of each individual spin (relative to
the position of the vacancy) was determined by the variational principle, namely (figuratively speaking)
by allowing for the delocalization of the hole without costing too much magnetic exchange 
energy. It is to be stressed that even though the exchange interactions are treated using classical spins, 
the vacancy is placed in a Bloch state and as it moves throughout the lattice the quantum--mechanical 
overlaps of the spins are fully accounted for.  A detailed account and summary of the results and 
predictions of the field theory that results from this work can be found in Ref. \cite {SSsummary}.

The extrapolation of this idea to the strong--coupling three--band Hamiltonian of Eq.~(\ref{eq:3tJ}) 
was carried out by Frenkel, one of the present authors, and Shraiman and Siggia \cite {frenkel}. 
However, when we applied the type of variational wave function developed in Ref. \cite {frenkel}
to the $E~ vs. \vec k$~ARPES data of Ref. \cite {wells}, we found that our fits led to ridiculous hopping 
energies \cite {nazarenko}.  Thus, we have produced a new variational wave function which we feel
better approximates the physics of the carrier motion of an Oxygen hole in a ${\rm CuO_2}$ plane.

To introduce our wave function it is instructive to first state the wave function that was
developed and studied in Ref. \cite {frenkel}. The Shraiman-Siggia--type wave functions incorporate
the broken symmetry of the AFM lattice; thus, referring to Fig. 1, for a model that includes
both the Cu and Oxygen sites there are two Cu basis sites (the spin up and spin down sublattices of the AFM)
and four Oxygen basis sites per primitive unit cell. 
Labeling the Cu sites by their position with respect
to the unit cell in which the Oxygen hole resides, {\em viz.}, the unit
cell denoted by $\vec r$,
one may write the wave function in the following form:
\begin{equation}
| \Psi_{\vec{k}} > = \sum_{\vec r} e^{i \vec{k} \cdot \vec{r}}~\sum_{\mu = 1}^{4} \gamma_\mu~
\psi_\mu^{O}(\vec{r})~\prod_{i} 
\psi_\mu (\vec{r}_{i}-\vec{r})~
\psi_\mu (\vec{r}_{i} + \hat x -\vec{r}) \quad,
\label{eq:3 band SC}
\end{equation}
In this equation, $i$ labels the primitive unit cells and $\vec r_i$ is 
the origin of
the $i$th unit cell. The product is over pairs of spinors, the pairs corresponding to the up
and down sublattice Cu spins in each unit cell, where, say, the up sublattice Cu site
is at $\vec r_i$ and then the down sublattice Cu site is at $\vec r_i + \hat x$.
The four basis sites at which the Oxygen hole may be found are denoted by $\mu = 1,2,3,4$ (see Fig. 1),
and the state of each Cu spinor can depend on which of the four sites the Oxygen hole is found at.
Finally, these states are put into Bloch states labeled by some wave vector $\vec k$.

The minimum energy state at each wave vector is found by minimizing the variational parameters
in this wave function. The variational parameters can be understood as follows: Consider 
one Oxygen hole location in a lattice with N unit cells.
There are 2N Cu sites implying that 4N angles fully specify the Cu spinors. 
Since there are four possible Oxygen hole locations per unit cell
and there is a complex amplitude $\gamma_{\mu}$ associated with each Oxygen position,
there are a total of 16N + 16 variational parameters.  However, since the absolute 
orientation of the spins and the overall phase of the wave function are arbitrary, 
there are only 16N + 12 variational parameters in this normalized
wave function that are physically meaningful to change. 
We have numerically obtained converged results as a function of increasing N.

As mentioned above, we previously found \cite {nazarenko} that the hopping parameters that are 
required to make the dispersion relation of this variational wave function best approximate 
the ARPES data of Ref. \cite {wells}, {\em viz.}, $t_a = 0.1~{\rm eV},
~t_b = 0.15$~eV, and $t_{pp} = -0.2$~eV, 
are physically unreasonable --- see  Eq.~(\ref{eq:hybert 3tJ energies}). Also,
the bandstructure produced from this wave function does not agree with the
experimentally observed flat band from $\vec k = (0,0) \rightarrow (\pi,0)$.
Thus, we now present the formalism and motivating ideas behind our new variational wave function.

One possible reason for the failure of the three--band semiclassical wave function
follows from the analysis of Zhang and Rice \cite {zhang_rice}: Quite simply,
these authors showed how local hopping processes tended to quench the spin of
the Oxygen hole to zero {\em at each Oxygen site}, thus producing spinless vacancies. 
In their formalism, this can only happen if the Oxygen hole is allowed to execute quantum 
fluctuations between its up and down spin states at each site --- clearly, 
Eq.~(\ref{eq:3 band SC}) does not include the possibility of such fluctuations.  
Assuming that the reduction of the Oxygen hole's spin is indeed physically
important (though not necessarily conceding that this spin is quenched to zero) we 
introduce the following class of variational wave functions.

We have generalized the variational wave function of Ref. \cite {frenkel} (presented
in Eq.~(\ref{eq:3 band SC})) by including Oxygen spin fluctuations. As we 
show below, the effect of this is to allow for quantum fluctuations of the
occupied Cu--O--Cu bond, thus allowing for the quasiparticle to approximate the
trion of Ref. \cite {ER}. The new variational wave function that we consider in
this paper is 
\begin{equation}
| \Psi_{\vec{k}} > = \sum_{\vec r} e^{i\vec k \cdot \vec r} 
\sum_{\sigma=\uparrow,\downarrow}~\sum_{\mu = 1}^{4} \gamma_{\mu,\sigma}
\psi_{\mu,\sigma}^{O}(\vec{r}) \prod_{i} \psi_{\mu,\sigma}(\vec{r}_{i}-\vec{r})
\psi_{\mu,\sigma}(\vec{r}_{i}+{\mathaccent 94 x}-\vec r)~~~~~,
\label{eq:vwf}
\end{equation}
where
\begin{eqnarray}
\psi_{\mu,\uparrow}^{O} = \left\{ \begin{array}{ll}
e^{-i \phi_{\mu}^{O}/2} & \cos (\theta_{\mu}^{O}/2) \\
e^{i \phi_{\mu}^{O}/2} & \sin (\theta_{\mu}^{O}/2)
\end{array}
\right\}
\label{eq:sup}
\end{eqnarray}
and
\begin{eqnarray}
\psi_{\mu,\downarrow}^{O} = \left\{ \begin{array}{ll}
e^{-i \phi_{\mu}^{O}/2} & \sin (\theta_{\mu}^{O}/2) \\
-e^{i \phi_{\mu}^{O}/2} & \cos (\theta_{\mu}^{O}/2)
\end{array}
\right\} .
\label{eq:sdown}
\end{eqnarray}
The angles specifying the Cu spinors $\psi_{\mu,\uparrow}$
and $\psi_{\mu,\downarrow}$ for the same Cu site are allowed
to be completely independent of one another.
However, by enforcing the structure of the spinors of Eqs.~(\ref{eq:sup},\ref{eq:sdown}) we ensure that
the $\sigma = \uparrow$ and $\sigma = \downarrow$ components of this variational wave function
are orthogonal to one another, and thus each product of spinors in this wave function is
orthogonal to every other product of spinors.

Armed with this wave function we attempted to find hopping parameters that were
close to those of Eq.~(\ref{eq:hybert 3tJ energies}) such that the measured
dispersion relation agreed with that produced by this new variational wave function.
Our results are shown in Fig. 2; clearly, the agreement that we find is superb. To the best of 
our knowledge, no other theory agrees with the ARPES data so accurately.

The energy parameters that enter Eq.~(\ref{eq:3tJ}) that correspond to our
fitted $E~vs.~\vec k$ data are 
\begin{equation}
J_{Cu-Cu} = 0.125~{\rm eV}, J_{Cu-O} = 0.63~{\rm eV}, t_a = 0.25~{\rm eV}, 
t_b = 0.3~{\rm eV}, {\rm and}~t_{pp} = -0.3~{\rm eV}~~~. 
\label{eq:final parameters}
\end{equation}
The comparison between the fitted hopping parameter values that we find and those 
obtained from perturbation theory/electronic structure work, {\em viz.} 
Eq.~(\ref{eq:hybert 3tJ energies}), is very pleasing except for the 
direct Oxygen--Oxygen hopping ($t_{pp}$) --- we find a value which is about half
of that predicted using other techniques. It is possible that the 10 \% larger lattice
constant of ${\rm Sr_2CuO_2Cl_2}$ in comparison to that used in the electronic
structure work is responsible for part of this difference.

In Fig. 2 we have also shown a solid curve which is a fit to the
energies obtained from our variational wave function.  The fit was done using
\begin{equation}
E(\vec{k}) = {\sum_{m,n}}^\prime A_{(m,n)} \cos^{m}(k_{x}) \cos^{n}(k_{y}),
\label{eq:parameterE}
\end{equation}
where $A_{(m,n)} = A_{(n,m)}$ and $m + n = 2N$ (N is an integer).  Only terms up
to and including N = 4 have been included in the calculation of this curve.
It is clear that Eq. (\ref{eq:parameterE}) preserves
the AFM symmetry ($E(\vec{k}) = E(\vec{k}+(\pm \pi, \pm \pi))$) and also the
reflections and rotations of the $C_{4v}$ point group.  The coefficients that
produce this curve are listed in Table I.

%  ** Physical wave function  **
%  -----------------------------
\section{Properties of the variational wave function}
\label{sec:physical_vwf}

In this section we discuss the properties of the variational wave functions that
were produced from fits to the ARPES data, and focus on two issues: (i) If the
one--band $t - t^\prime - J$ model does not lead to a fit with ARPES whereas
the three--band model does, what's the problem with the one--band model? Here we 
display results showing that the spin of the Oxygen hole is not quenched
to zero, and thus the mobile carrier does not have a spin that is quenched
to zero as it is (by definition) for the spinless vacancies of the one--band 
$t - t^\prime - J$ model. (ii) If the three--band to one--band (spinless vacancy)
renormalization is not adequate for all wave vectors in the first Brillouin zone, 
then Zhang--Rice quasiparticles are not a good description of the carriers  everywhere in
the first Brillouin zone --- so, what is a good description of
the quasiparticles? Here we display our quantitative data verifying that the
trion quasiparticle of Ref. \cite {ER} is an excellent representation of
mobile carriers.

To indirectly evaluate the validity of the one--band model in describing the
quasiparticles propagating in ${\rm Sr_2CuO_2Cl_2}$, we have determined
the magnitude of the spin of the Oxygen hole as a function of wave vector
using the variational wave functions determined in the last section.
To this end, we have computed the expectation value of the Oxygen spin operator:
\begin{equation}
< \vec{S}^{O}(\vec{k}) > = \sum_{\mu} < \Psi_{\vec{k}} | \hat {\vec{S}^{0}_{\mu}}|
\Psi_{\vec{k}} >,
\label{eq:oxygenspin}
\end{equation}
where the sum over $\mu$ refers to the four Oxygen basis sites.
Our results for the magnitude of the quantity defined in Eq. (\ref{eq:oxygenspin}) 
are shown in Fig. 3.  We see that along the AFM Brillouin 
zone the Oxygen hole's spin is essentially quenched to zero and thus we
would expect the one--band model to be a good representation of the
quasiparticles along the AFM Brillouin zone --- the one--band $t - t^\prime - J$ model
of Ref. \cite {nazarenko} indeed agrees with the ARPES data along this branch.   
However, along the other high--symmetry branches of the first Brillouin zone 
we find that the spin of the Oxygen hole is not quenched to zero and is in fact
quite large. Concomitantly, it is along $(0,0) \rightarrow (\pi,0)$ that
the one--band $t - t^\prime - J$ model has difficulties fitting the ARPES data.  
Thus, if these three--band variational wave functions are indeed a good 
representation of the true wave functions of this system, any renormalization that 
attempts to quench this spin and describe the motion in terms of spinless vacancies 
is not necessarily justified and not necessarily going to work.

If the spin of the carrier is not quenched to zero, then the spinless vacancy
quasiparticles which are the biproduct of the Zhang--Rice renormalization \cite {zhang_rice}
cannot be adequate. The question then remains: what is the best representation
of the quasiparticles?  An answer first put forward by Emery and Reiter,
{\em viz.} so--called trions, can
best be understood by considering the case of a single Oxygen hole localized
on a Cu--O--Cu bond. From Eq.~(\ref{eq:3tJ}) we have that the Hamiltonian
for this bond is nothing more than
\begin{equation}
H = J_{Cu-O}~\vec{S}^{O} \cdot (\vec{S}_1 + \vec{S}_2)
\label{eq:3siteH}
\end{equation}
where $\vec{S}_1$ and $\vec{S}_2$ are the respective spins of the two Cu spins
that are neighbours
to the Oxygen hole. Letting the components of each ket refer to a 
Cu$_1$-O-Cu$_2$ labeling 
of the sites, the ground state of
this problem is a doublet, with the $S^z_{Total} = 1/2$ wave function being given by
\begin{equation}
|\Psi_g, \uparrow> = {1\over \sqrt{6}}~( 2 |\uparrow, \downarrow, \uparrow> - |\uparrow, \uparrow,
\downarrow> - |\downarrow, \uparrow, \uparrow>)~~~~.
\label{eq:trion}
\end{equation}
Emery and Reiter \cite {ER} surmised that this and its time--reversed state were the quasiparticles, so--called
trions, that moved through the lattice when the Oxygen hole was delocalized. In fact, 
if the spin distortions of the AFM background were prohibited (unlike our Eq.~(\ref{eq:vwf})), 
these quasiparticles were found by Klein and Aharony \cite {klein} to best describe carrier motion 
in ${\rm CuO_2}$ planes.

As discussed in Ref. \cite {frenkel},
an excellent representation of the true ground state of this Hamiltonian follows from a 3--spin
version of the variational wave function given in Eq.~(\ref{eq:vwf}) of this
paper. In fact, the success of this kind of problem in solving the above Hamiltonian
was one of the motivations for our consideration of Eq.~(\ref{eq:vwf}) --- to be specific,
if we could ably represent the trions and they were indeed the appropriate quasiparticles
of the mobile hole problem, then these should be excellent variational wave functions.
We formed the density matrix for the possible states of the occupied Cu--O--Cu bond
and calculated the trion occupation for our variational wave functions found from
fitting to the ARPES data; our results are shown in Fig. 4. Independent of the wave vector 
we find that our wave functions have an occupation of at least 92 \% in the trion states.
Note that for the high symmetry wave vectors the occupation is between 97 per
cent and 98 per cent.  Therefore, we find that the trion doublet state is an excellent
representation of the quasiparticle for these planes. 

%  ** Discussion **
%  ----------------
\section{Discussion}
\label{sec:discussion}

We have presented our formalism for the extension of the Shraiman--Siggia 
semiclassical variational wave functions to the strong--coupling three-band 
model including the quantum fluctuations of the occupied Cu-O-Cu bond.
For reasonable hopping parameters we find superb agreement with the $E~vs.~\vec k$
dispersion relation that is measured experimentally in antiferromagnetically
aligned ${\rm Sr_2CuO_2Cl_2}$. Since it was found that the single--band 
$t - t^\prime - J$ model was not able to fit the quasiparticle dispersion 
throughout the entire first Brillouin zone, perhaps the three-band model discussed 
in this paper will be necessary to describe the normal--state properties of these planes. 
However, we can not determine the full spectral function using this approach, and thus
a more complete comparison of the measured $A(\vec k, \omega)$ to our
three-band model is still lacking.

Theories of related spectral functions do indeed exist, but are unfortunately
inadequate to treat the task at hand:
One of the most popular analytical theories of the $t - J$ model is the
self--consistent Born approximation (SCBA) \cite {marsiglio,martinez,liu}. It was
argued in Ref. \cite {liu} that vertex corrections of the so--called spin polaron model 
version of the $t - J$ model were absent ({\em viz}, those of the two--loop crossing
diagram) to second--order in perturbation theory, and thus it was expected that this theory 
should be very successful.  This forecast was indeed verified in a recent {\em exact}, 
unbiased numerical determination of the spectral function for a single hole propagating 
in the largest square cluster yet doped, {\em viz.} a 32--site cluster, as described 
by the $t - J$ model \cite {32site} --- the comparison 
of the exact $E~vs.~\vec k$ dispersion relation to that predicted by the SCBA led to an 
impressive agreement. 

Unfortunately, if one includes either a next--nearest--neighbour hopping $t^\prime$ in 
the one--band model, or applies this same formalism to the three-band model \cite {reiter},
the same cancellation of low--order vertex corrections does not occur. Consequently,
such approximate theories are much more speculative. Thus, vertex
corrections must be examined and understood before an analytical theory of the spectral
function is available, and only when it is available will a full comparison of the
spectral function of Eq.~(\ref{eq:3tJ}) to that of the experimental data of Ref. \cite {wells}
be possible.

These new wave functions, and the associated $E~vs.~\vec k$ dispersion
relation, can be applied to the calculation of quantities such as the Pauli susceptibility 
and specific heat using the rigid--band approximation --- these results 
will be presented in a planned future publication.

\newpage
%  ** Acknowledgments**
%  --------------------
\centerline{Acknowledgments:}

We wish to thank Barry Wells, Z.--X. Shen and Bob Birgeneau for informative discussions regarding their
${\rm Sr_2CuO_2Cl_2}$ ARPES data. Also, we wish to acknowledge fruitful discussions on the application of
the one--band model to the ARPES problem with Alexander Nazarenko, Stephan Haas and Elbio Dagotto.
We thank George Reiter for sending us his preprint prior to publication.
This work was supported by the NSERC of Canada.

\bigskip
\noindent

%  ** References **
%  ----------------

%
% Figure Captions
%
\newpage
{\bf Figure Captions}
\begin{enumerate}

\item
\label{fig:orbitals}
A schematic of the two Cu $d_{x^2 - y^2}$ and four Oxygen $p_x/p_y$ orbitals
of the primitive unit cell of a state with broken AFM symmetry.

\item
\label{fig:energy}
Quasiparticle energies (solid diamonds) of an Oxygen hole in the effective
Hamiltonian of Eq. (\protect\ref{eq:3tJ}) for an infinite lattice using our new 
variational wave function that includes the spin fluctuations of the Oxygen
carriers.  The quasiparticle dispersion measured in Ref. \protect\onlinecite {wells} is also
shown (open circles).  The parameters that produce this quasiparticle dispersion
are $J_{Cu-Cu} = 0.125~{\rm eV}, J_{Cu-O} = 0.63~{\rm eV}, t_a = 0.25~{\rm eV}, 
t_b =0.3~{\rm eV}$, and $t_{pp} = -0.3~{\rm eV}$.

\item
\label{fig:oxygenspin}
The magnitude of the Oxygen spin operator's expectation value (solid diamonds)
for a single Oxygen hole on an infinite lattice (c.f.
Eq.~(\protect\ref{eq:oxygenspin})).

\item
\label{fig:trion}
The percentage occupation of the spin--up and spin--down trion doublet states
in our variational wave functions as a function of wave vector.
\newpage
\begin{table}
\caption{Coefficients appearing in Eq. (8) that describes
the $E~vs.~\vec k$ relation predicted by our variational wave
function for the parameters of Eq. (7).}
\vskip 0.7 truecm
\begin{tabular}{ccccc}
&$i$&$j$&$A_{ij}$ (eV)\\
\tableline
& 2 & 0 & 0.1755 \\
& 1 & 1 & 0.03078 \\
& 4 & 0 & 0.6628 \\
& 3 & 1 & 0.1520 \\
& 2 & 2 & 0.2694 \\
& 6 & 0 & -0.7557 \\
& 5 & 1 & -0.3868 \\
& 4 & 2 & -0.7220 \\
& 3 & 3 & -0.1955 \\
& 8 & 0 & 0.2190 \\
& 7 & 1 & 0.1484 \\
& 6 & 2 & 0.3315 \\
& 5 & 3 & 0.1420 \\
& 4 & 4 & 0.2246 \\
\end{tabular}
\end{table}

\end{enumerate}

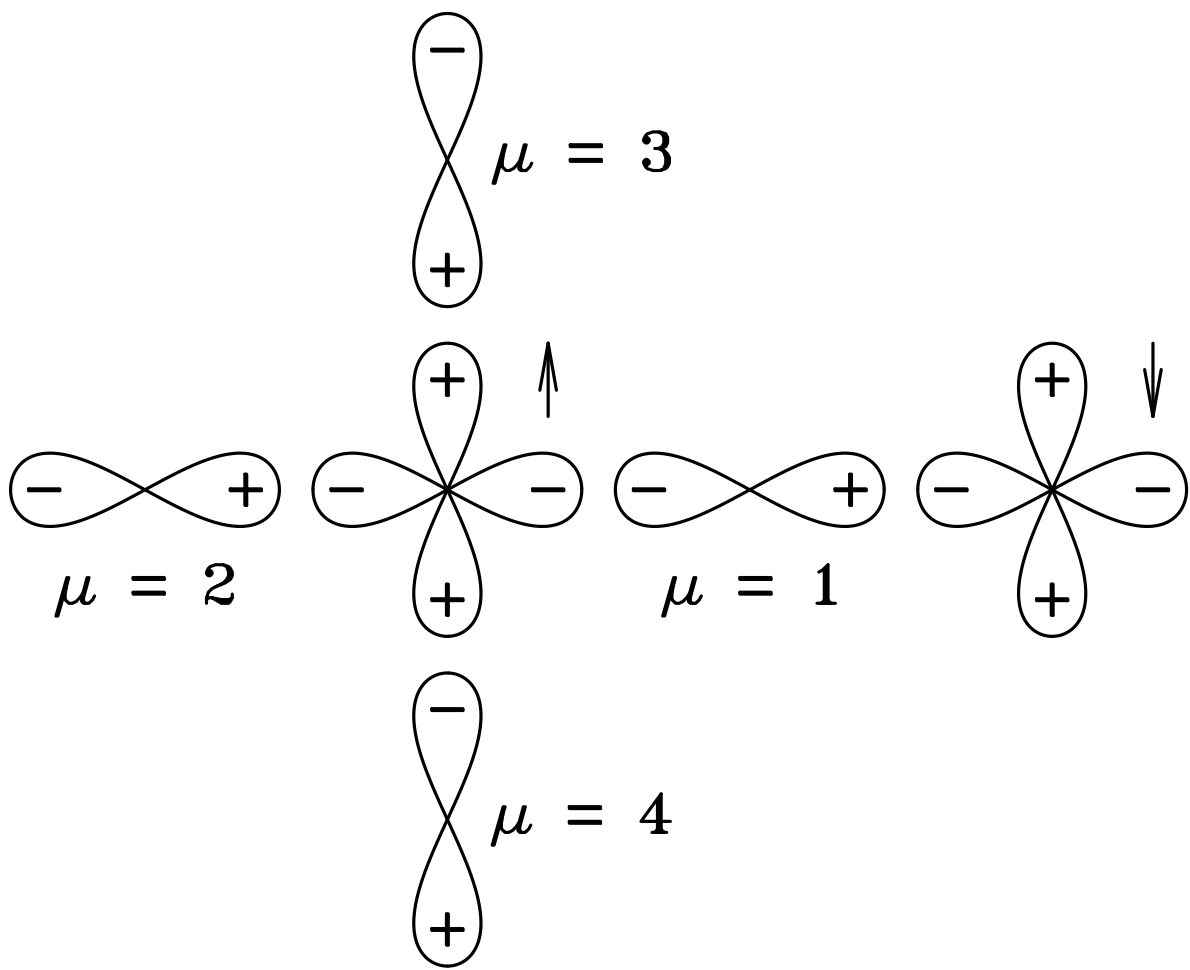
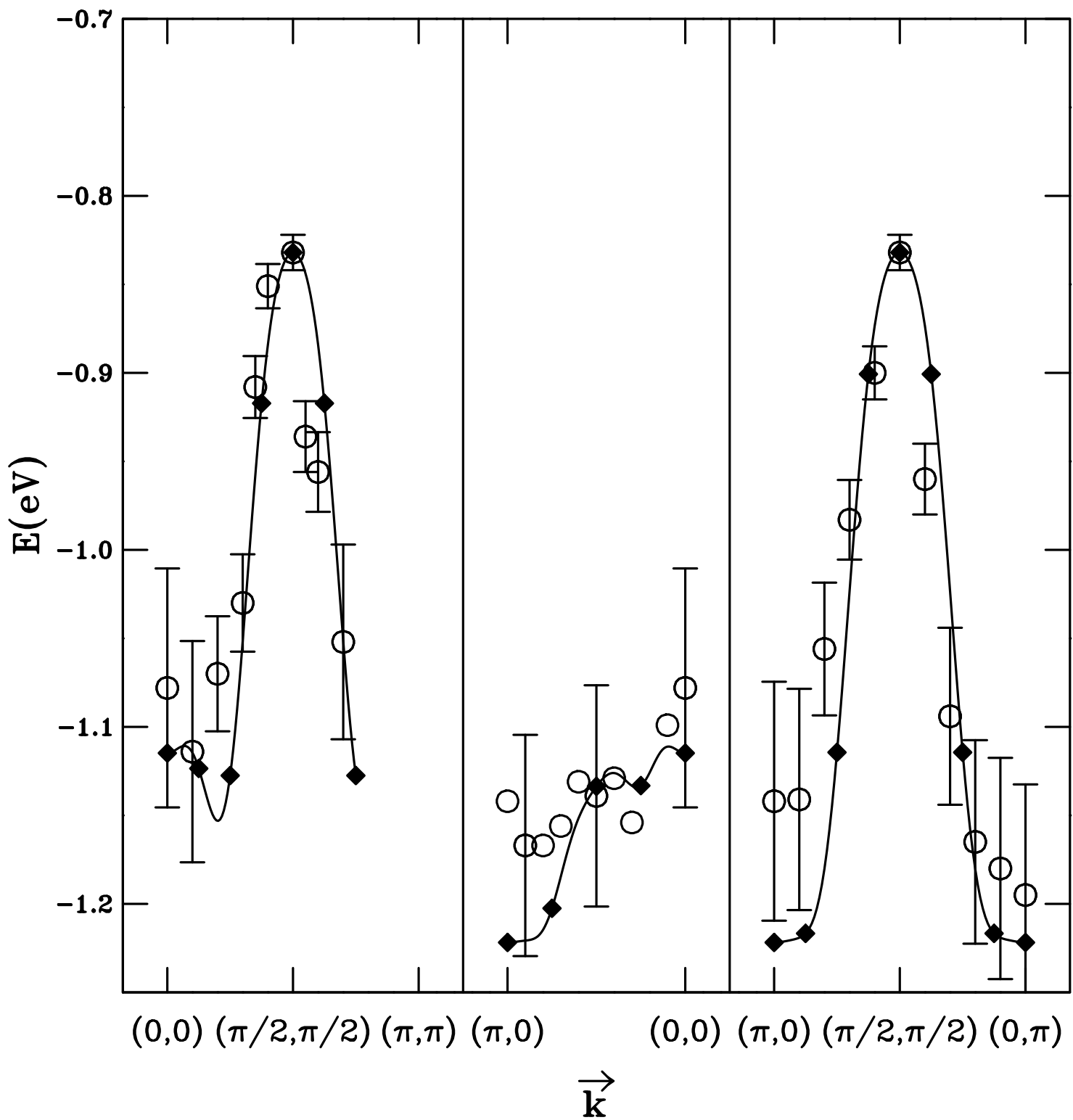
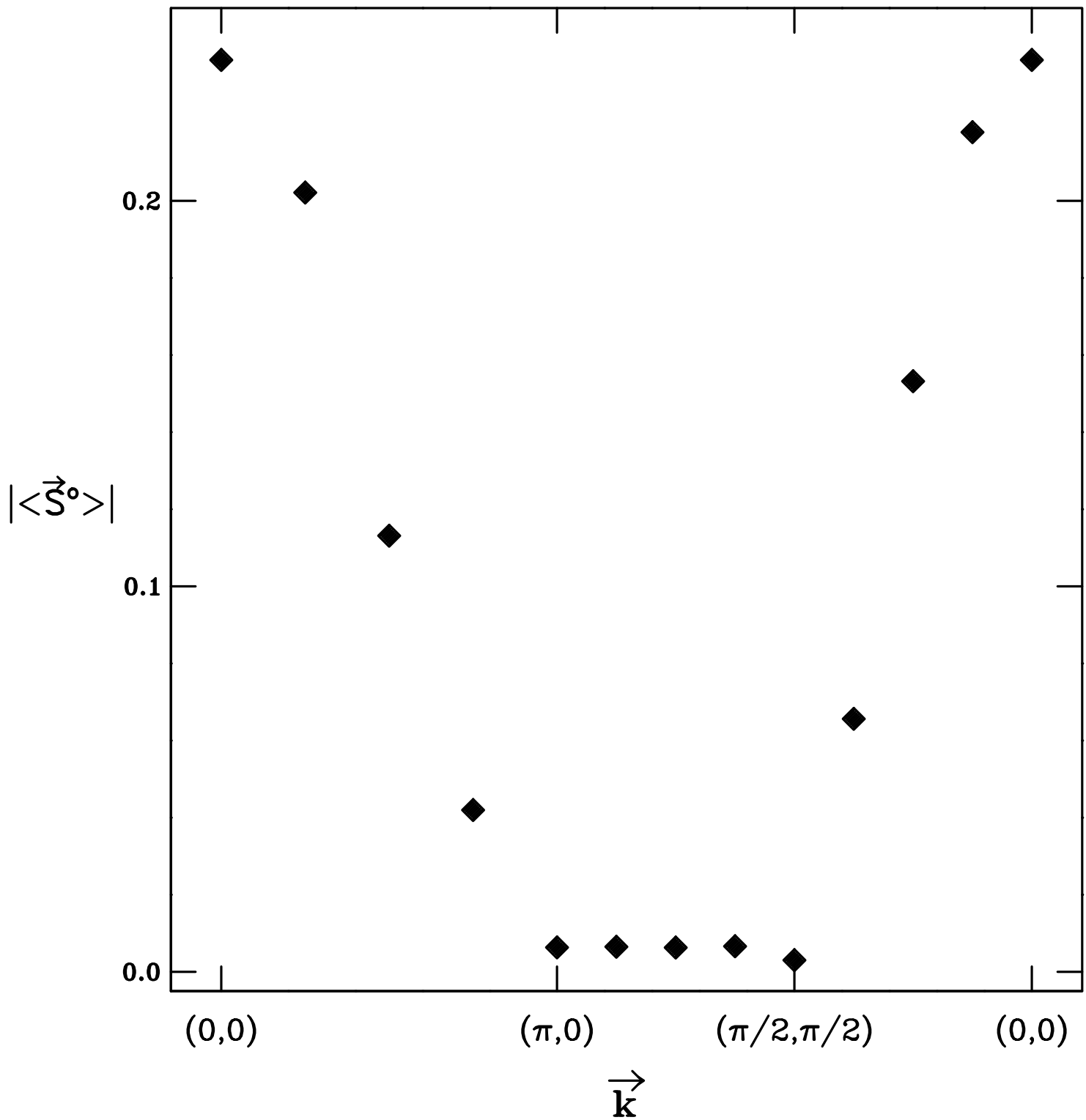
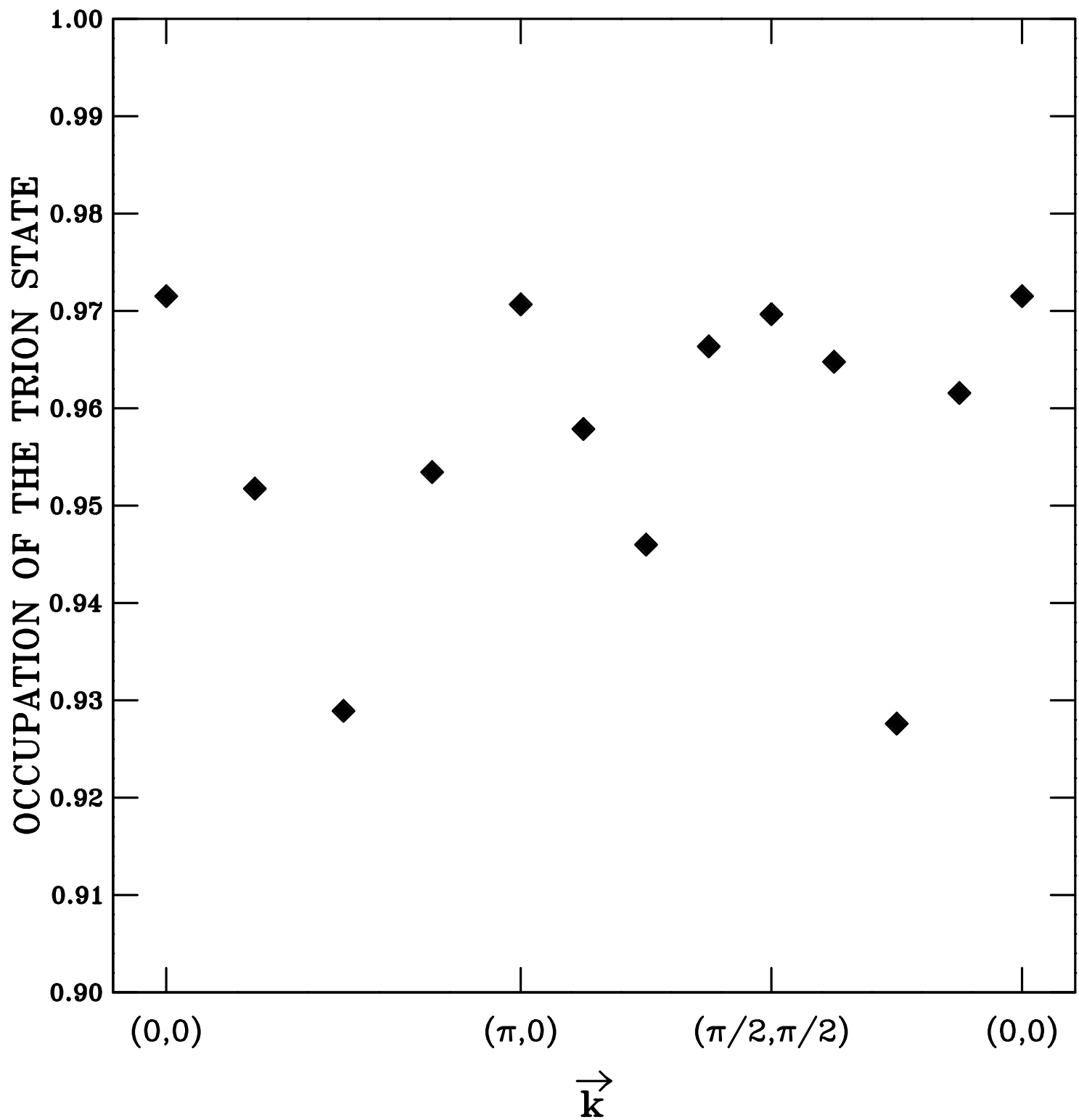
\end{document}